\documentclass[apjl]{emulateapj}

\usepackage{graphicx}
\usepackage{subfigure}

\usepackage{natbib}
\begin{document}


\title{Solar neutrino physics oscillations: \\
Sensitivity to the electronic density in the Sun's core}

\author{Il\'idio Lopes\altaffilmark{1,2,4} and Sylvaine Turck-Chi\`eze\altaffilmark{3,5}}

\altaffiltext{1}{Centro Multidisciplinar de Astrof\'{\i}sica, Instituto Superior T\'ecnico, 
Universidade Tecnica de Lisboa , Av. Rovisco Pais, 1049-001 Lisboa, Portugal} 
\altaffiltext{2}{Departamento de F\'\i sica,Escola de Ciencia e Tecnologia, 
Universidade de \'Evora, Col\'egio Luis Ant\'onio Verney, 7002-554 \'Evora - Portugal} 

\altaffiltext{3}{CEA/IRFU/Service d'Astrophysique, CE Saclay, 91191 Gif sur Yvette, France} 
\altaffiltext{4}{E-mail:ilidio.lopes@ist.utl.pt;ilopes@uevora.pt} 
\altaffiltext{5}{E-mail:sylvaine.turck-chieze@cea.fr}

\begin{abstract}
Solar neutrinos coming from different nuclear reactions are now detected with a high statistics. 
Consequently,  an accurate spectroscopic analysis of the neutrino fluxes arriving on the Earth's detectors become available, in the context of neutrino oscillations. In this work, we explore the possibility of using this information  
to infer the radial profile of the electronic density in the solar core. So,  we discuss the constraints on the Sun's density and chemical composition 
that can be determined from solar neutrino observations. 
This approach  constitutes an independent and 
alternative diagnostic to the helioseismic investigations already done.   
The direct inversion method, that we propose to get the radial solar electronic density  
profile, is  almost independent  of the solar model. 
\end{abstract}

\keywords{Neutrinos -- Sun:evolution --Sun:interior -- Stars: evolution --Stars:interiors}

\maketitle

\section{Introduction\label{sec-intro}}

Neutrinos, once produced in the core of the Sun, reach the Earth in less than 8 minutes.
On their  interplanetary journey, solar neutrinos  very rarely interact with other particles.
Luckily for physicists, 
neutrinos with a scattering cross-section varying from some  $\rm 10 ^ {-46} cm ^ 2 $ to 
some $\rm 10 ^ {-42} cm ^ 2 $~\citep{1989neas.book.....B}, occasionally  
interact  with  the baryons present in the solar neutrino detectors.
Therefore, the neutrinos produced in the nuclear reactions of  proton-proton (PP) chains 
and Carbon-Nitrogen-Oxygen (CNO) cycle are natural probe-particles of the constitutive
 plasma of the Sun's core.
 
The current theory of neutrino physics, following in the footsteps of the fundamental theory of elementary particles, has shown that
neutrinos  appear in three (at least)  flavours \citep{2010arXiv1010.4131H}: electron-neutrino ($\nu_e$),  muon-neutrino
($\nu_\mu$) and tau-neutrino ($\nu_\tau$).  The theory states that by 
vacuum flavour oscillations,  when moving in vacuum,  neutrinos have the ability 
to switch cyclically between different flavours.  This  mechanism of neutrino oscillations was proposed by Pontecorvo around 1940.  
In the late seventies, Lincoln Wolfenstein and colleagues suggested  another mechanism for neutrinos to change flavour, 
as neutrinos move through a high density medium \citep{1978PhRvD..17.2369W}: in an environment of high density material, 
the effective mass of the propagating neutrinos is different from the mass of neutrinos that propagates in a vacuum.
Since the oscillations between different flavours of neutrinos depend on their mass, the oscillations of neutrinos 
in dense media are different from neutrino oscillations in vacuum. This  process is now known as 
Mikheyev-Smirnov-Wolfenstein (MSW)  or "matter oscillations".
These two mechanisms of flavour oscillations are central to the modern theory of neutrinos.
 Through these mechanisms the electron-neutrinos produced in the Sun's core have a non-zero probability  
of being detected as a muon-neutrino or tau-neutrino when they reach the Earth. 
In particular  a small fraction of electron-neutrinos changes  flavour  in  the Sun's interior
when they propagate through  the high density plasma medium of the Sun's core. 
 
In recent years, significant progress has been achieved in the understanding of 
the mechanism responsible for the neutrino oscillations \citep{2003PhRvL..90b1802E,2001PhRvL..86.5651F,2001PhRvL..87g1301A}. 
The current theory of neutrino physics successfully explains the  neutrino observations.
Furthermore, the theory has been used successfully  to determine the  
fundamental properties of  neutrinos \citep{2008NJPh...10k3011S}. 

If, during several decades, the solar neutrino fluxes arriving on Earth  have been difficult to interpret, this is no longer the case as the flavour oscillations at different energies are well understood. In particular, the production of these neutrinos has been confirmed by helioseismic data. One may consider as a success that these two disciplines (solar neutrinos and helioseismology)  agree within the error bars 
\citep[see][and references therein]{2010ApJ...715.1539T, 2011RPPh...74h6901T}. 
Such an agreement, will now allows us to progress a step further.
We believe that solar neutrinos will very likely become  a powerful tool
to probe the core of the Sun because the production of neutrinos is very sensitive 
to the local properties of the solar plasma. 

In this paper, we focus on the sensitivity of solar neutrinos to the 
plasma in the solar core, and explore the possibilities of probing the physical properties of
such plasma by using the solar neutrino flux measurements.
This point has been proposed by 
John Bahcall and Raymond Davis about 50 years ago  \citep{1964PhRvL..12..303D}.
In principle, the structure of the solar core can be studied by means of neutrino spectroscopy
in two fundamental ways: (i) to diagnose the temperature profile in the Sun's core, 
by measuring the total number of electron-neutrinos produced in  each nuclear reaction of PP
chains and CNO cycle, in particular the boron flux that is the most sensitive to the temperature \citep{2012ApJ...746L..12T, 2012RAA...12h1107T}
and (ii) to measure the radial electronic density of matter of the Sun
by determining the amount of electron-neutrinos that are converted into another flavour. 
This article is a first attempt to study  this second physical information through neutrino probes.
 
The neutrino flavour oscillations by the MSW mechanism are particularly significant in the solar interior, 
where there is a wide radial variation of the plasma density.  
The standard solar model \citep{1993ApJ...408..347T}  partly validated by helioseismology,
predicts that the density inside the Sun varies
from  about 150 ${\rm g\;cm^{-3}}$ in the centre of the star, to  1 ${\rm g\;cm^{-3} }$ at half of the solar radius.

What makes this diagnostic  particularly powerful is the 
possibility of getting a direct measurement of  the plasma electronic density almost independently of the solar model.
The strong dependence of neutrino oscillations with the local electronic density of matter opens the possibility 
of inferring the matter density and chemical composition in the nuclear region.  Since the neutrino oscillations in matter depend strongly on the matter density  
and  weakly on the chemical composition,  this is at first, a diagnosis of the radial density profile 
of matter in the solar core.  The composition diagnostic can come from  precise CNO neutrino detections. Therefore, we can anticipate that with the expected increased accuracy  of the solar neutrino detections in the future, it will be possible to extend this analysis to the solar core chemical 
composition determination.

\begin{figure}[!t]
\centering 
\includegraphics[scale=0.50]{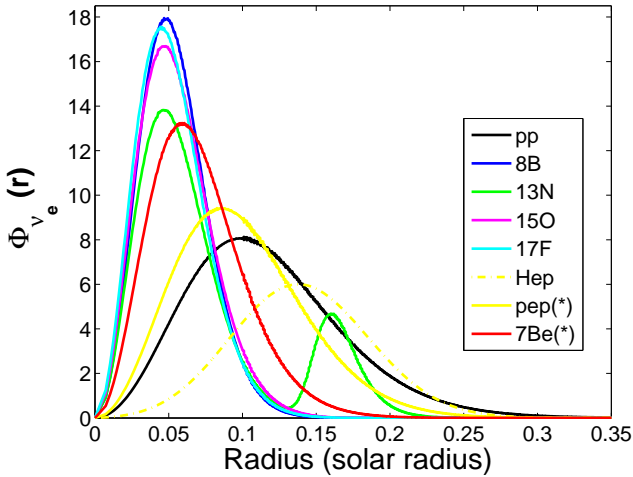}
\caption{The electron-neutrino fluxes produced in the various nuclear reactions of the  
PP chains and CNO cycle.  
The center of the different neutrino sources $\Phi_j (r)$,  $\langle r \rangle_j $  
(with $j=pep,pp,^8B,^7Be,^{13}N,^{15}O,^{17}F$) are the following ones (in unity of $R_\odot$):
0.099 ($pep$), 0.117 ($pp$),0.125 ($Hep$), 0.054 ($^8B$), 0.070 ($^7Be$), 0.074 ($^{13}N$), 0.055($^{15}O$), 0.053($^{17}F$).
These neutrino fluxes were calculated for the standard solar model, using the most updated microscopic physics data.
This solar model is in agreement with the most current  helioseismology diagnostic and other solar standard models 
published in the literature (see text).
For each neutrino type $j$, $\Phi_j (r)\equiv (1/F_j)\;d{f_j(r)}/dr$ is drawn
as a function of the fractional radius $r$ for which $f_j$ is the flux in ${\rm s^{-1}}$ 
and $F_j$ is the total flux for this neutrino type.
}
\label{fig:A}
\end{figure}

\section{Current Model of neutrino physics oscillations}

The theory of neutrino oscillations in vacuum and matter \citep{2010arXiv1010.4131H,2008NJPh...10k3011S} 
has successfully addressed the problem of the solar neutrino deficit - the discrepancy between the neutrino flux detected 
and the theoretical predictions of the solar standard model \citep{1993ApJ...408..347T,2009arXiv0905.3549F,2008PhR...460....1G}. 
The theory  provides a theoretical solution in full agreement with all solar neutrino experiments 
\citep{2011PhLB..696..191B,2010PhRvD..82c3006B,2010PhRvC..81e5504A} as well as with the data 
obtained  by the KamLAND reactor experiment \citep{2003PhRvL..90b1802E}.
Presently, the theoretical model for neutrino flavour oscillations  is defined by means of six parameters: 
the difference of the squared characteristic masses  ($ \Delta m^2_{12}, \Delta m^2_{31} $),
 the mixing angles  ($\theta_{12}, \theta_{23}, \theta_{31} $) and the CP-violation phase. 
The mass square differences and mixing angles are known 
with reasonable accuracy \citep{2009arXiv0905.3549F}:  $\Delta m_{31}^2 $ is obtained from the experiments 
of atmospheric neutrinos and $\Delta m_{12}^2 $  is obtained from solar neutrino experiments. 

 The mixing angles are not  uniformly well defined: $\theta_{12} $ is
 obtained from solar neutrino experiments with an excellent precision; 
 $\theta_{23} $ is obtained from the atmospheric neutrino experiments, 
 this is the mixing angle of the highest value; $ \theta_{13} $ has been first estimated from the 
 Chooz reactor \citep{1999PhLB..466..415A}, its value is very small and 
 was still  very uncertain \citep{2009arXiv0905.3549F}.  Nowadays with Daya Bay and Reno, the situation is largely improved \citep{2012PhRvD..86a3012F}, as shown in the next paragraph. 
 However,  present experiments cannot fix 
 the value of  the CP-violation phase \citep{2010arXiv1010.4131H}. 

An overall fit to the data obtained from the different  neutrino experiments: 
solar neutrino detectors, accelerators, atmospheric neutrino detectors 
and nuclear reactor experiments suggests that  the parameters of neutrino oscillations are the following ones 
\citep{2008PhR...460....1G}: $ \Delta m_{31}^ 2\sim 2.46 \pm 0.12\; 10^{-3} eV^2 $, 
$ \Delta m_{12}^2 \sim 7.59 \pm 0.20 10^{-5} eV^2$, 
$ \theta_{12} = (34.4 \pm 1)^o, \theta_{23} = (42.8_{-2.9}^{+4.7})^o $ and $ \theta_{13} = (5.6_{-2.7}^{+3.0} )^o $.  The recent progress leads to $ \theta_{13} = (8.6_{-0.46}^{+0.44} )^o$ and  $ \Delta m_{32}^2 \sim -2.43 _{-0.06}^{+0.42} 10^{-3} eV^2$\citep{Gonzalez}.

In the limiting case where the value of the  mass differences, $ \Delta m^2_{12}$ or $ \Delta m^2_{31}$ is large, 
or one of the angles of mixing ($ \theta_{12}, \theta_ {23}, \theta_{31} $) is small, 
the theory of three neutrino flavour oscillations reverts to an effective theory of two neutrino flavour oscillations
\citep{2010arXiv1010.4131H}. 
Balantekin and Yuksel have shown that the survival probability of solar neutrinos 
calculated in a model with two neutrino flavour oscillations or three neutrino flavour oscillations have very close values \citep{Balantekin:2003vi}.

In the present work,  for reasons of convenience and simplicity,
we will restrict our study to the theory of two neutrino flavour oscillations. 
The generalization of the results to a theory of three neutrino flavour oscillations is obvious
\citep{1997PhRvD..56.1792L,2011PhRvD..83e2002G}.  
The  survival probability of electron-neutrino function  $P_{\nu_e} (\equiv P (\nu_e \rightarrow \nu_e) $ )
in  a two flavour neutrino theory  is given by 
\begin {eqnarray} 
P_{\nu_e}(E, r) = 
\frac {1} {2} + F (\gamma) \cos {(2 \theta_v)} \cos{(2 \theta_m)} 
\label{eq-A}
\end {eqnarray} 
where $ \theta_v $ is the angle of oscillation in vacuum and $ \theta_m$ is the angle oscillation in matter. 
 The function $ F (\gamma) $ is a first-order correction to the adiabatic approximation of the propagation 
of neutrinos in matter, where $\gamma$ is the adiabatic parameter which depends of the local properties of the plasma~\citep{1932ZPhy...78..847L}.  The adiabaticity occurs when the electronic density of the propagating medium is a slow varying function  over the neutrino travel path (or equivalently the solar radius).  
This approximation is valid in the solar interior, in the case where 
$ \gamma \gg 1 $,  a condition that is verified in much of the solar core and  the radiative region \citep{1932ZPhy...78..847L,1932RSPSA.137..696Z}.  As $ F (\gamma) =0.5-P_{12}$ where $P_{12}$ corresponds to the transition between two distinct neutrino flavours, at first order it is reasonable to consider that $P_{12}=0$. The numerical value of $ F (\gamma) $ is approximately equal to 0.5 \citep{Bilenky:2010uf}. 
The angle  $ \theta_v$ depends on the distance from the Sun to the Earth. In this model we assume $ \theta_v$ 
to be equal to $ \theta_{12}$. The phase $ \theta_m $ is the most important term in this analysis, 
since it depends on the properties of the plasma in the Sun's core, namely, the 
 local electron density.  The neutrino mixing angle  $\theta_m$  is given by 
\begin{eqnarray} 
\sin{2\theta_m}= 
\frac{\sin{2\theta_v}}
{\sqrt{(V/\Delta m^2-\cos{(2\theta_v)})^2+ \sin^2{(2\theta_v)}} } 
\label{eq-B}  
\end{eqnarray}
where $ \Delta m^2 $ is the difference of the squared masses we consider to be equal to $ \Delta m_{12}^2 $. 
The function of the solar plasma $V $ is given by
\begin{eqnarray} 
V (E,r) = 2 \sqrt{2} G_f \; n_e (r) E
\label{eq-B2}  
\end{eqnarray}
where $ G_f $ is the Fermi constant, $ n_e (r) $ is the electron density of plasma and $ E $ 
the energy of the neutrino. The electron density  $ n_e (r) =N_{o}\; \rho(r)/\mu_e(r) $ 
where $\mu_e$ in the mean molecular weight per electron, $\rho(r)$ the density of matter 
and $N_{o}$ the Avogadro's number.

\section{Neutrino production in the Sun's core}

The neutrino fluxes produced in the various nuclear reactions of the PP chains and CNO cycle have been 
computed for an updated version of the solar standard model 
\citep{turck04,2005ApJ...618.1049B,2010ApJ...713.1108G,2010ApJ...715.1539T,2011RPPh...74h6901T}. 
Figure~\ref{fig:A} shows the location of the different neutrino emission regions of the nuclear reactions. 
In the Sun's core, the neutrino emission regions occur in a sequence of shells, following closely  
the location of nuclear reactions,  orderly arranged in a sequence 
dependent on their temperature.  

The first reaction of the PP chains, the $pp$ nuclear reaction, has the largest neutrino emission shell. 
This region extends from the center  of the Sun  up to 30 \% of the solar radius. The $pep$ reaction has 
a neutrino emission shell identical to the $pp$ reaction, although only up to 25 \% of the solar radius.  
These two key nuclear reactions  are strongly dependent  on the total luminosity of the star. 
This is the reason why different solar models with the same total luminosity produced the same 
$pp$ and $pep$ neutrino emission shells. The neutrino emission shells
of $^8B$-$\nu$ and $^7Be$-$\nu$ extend up to  15 \% and 22 \% of the solar radius. 
Finally, it is worth noticing that the maximum emission of neutrinos for the PP chains
nuclear reactions, follows an ordered sequence  (see figure~\ref{fig:A}): $^8B$-$\nu$,
 $^7Be$-$\nu$, $pep$-$\nu $ and $pp$-$\nu$  
with the maximum emission located at 5 \%, 6 \%, 8 \% and 10 \% of the solar radius, respectively.

The CNO cycle nuclear reactions produce the following electron-neutrinos sub-species: 
$^{15}O$-$\nu$, $^{17}F$-$\nu$ and $^{13}N$-$\nu$
released in  emission  shells 
identical to the $^8B$-$\nu$. The $^{13}N$-$\nu $  have a second emission 
shell located between 12\% and 25 \% of the Sun's radius. The emission  of neutrinos 
for $^{15}O$-$\nu$, $^{17}F$-$\nu$ and $^{13}N$-$\nu$ is maximal at 4\%-5 \%  of the solar radius. 
The $^{13}N$-$\nu$ neutrinos have a second emission maximum which is located at 16 \% of the solar radius.

\section{Neutrino Flavour Oscillation in the Sun}

\begin{figure}[!t]
\centering 
\includegraphics[scale=0.50]{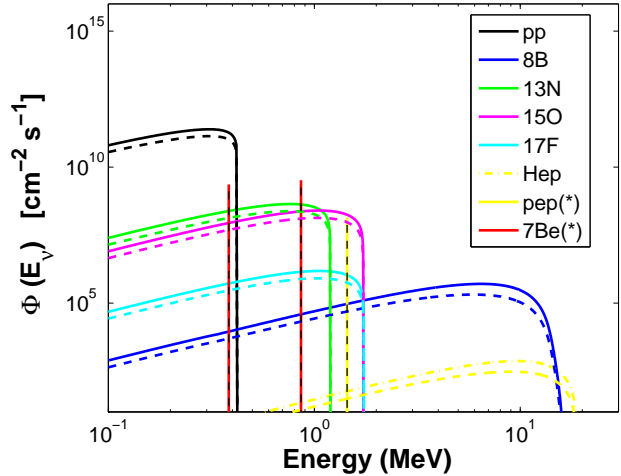}
 \caption{
The solar neutrino energy spectrum predicted by the standard solar model. 
 The solid curves correspond to the  total  neutrino fluxes  produced in the various nuclear reactions of the  PP chain and CNO cycle. The dashed curves correspond to  electron-neutrino fluxes of the various nuclear reactions after neutrino flavour conversion. 
The neutrino fluxes from continuum nuclear sources are given in units
of  cm$\rm ^{-2}s^{-1}Mev^{-1}$.
The line fluxes (indicated in the legend with $(*)$) are given in cm$\rm ^{-2}s^{-1}$. }
\label{fig:B}
\end{figure}

The neutrino emission reactions of  the PP chains and the CNO cycle are produced at high temperatures
in distinct layers in the Sun's core. Similarly, the neutrino flavour oscillations occur in the same 
regions. The average survival probability of electron-neutrinos in each nuclear reaction region is given by
\begin{eqnarray} 
\langle P_{\nu_e } (E)\rangle_j = 
A_j^{-1} \int P_{\nu_e} (E,r)\Phi_j (r) 4\pi \rho(r) r^2 dr 
\label{eq-C}
\end{eqnarray}  
 where $ A_j $ is a normalization constant given by  $ A_j= \int \Phi_j (r) 4 \pi \rho (r) r^ 2  \;dr $ 
 and $ \Phi_j (r) $  is the electron-neutrino emission function for the $j$ nuclear reaction. $j$ corresponds 
 to the following electron-neutrino reaction subspecies: $pp$, $pep$, $^8B$, $^7Be$, $^{13}N$, $^{15}O$ and $^{17}F $.
 
\begin{figure}[!t]
\centering 
\includegraphics[scale=0.30]{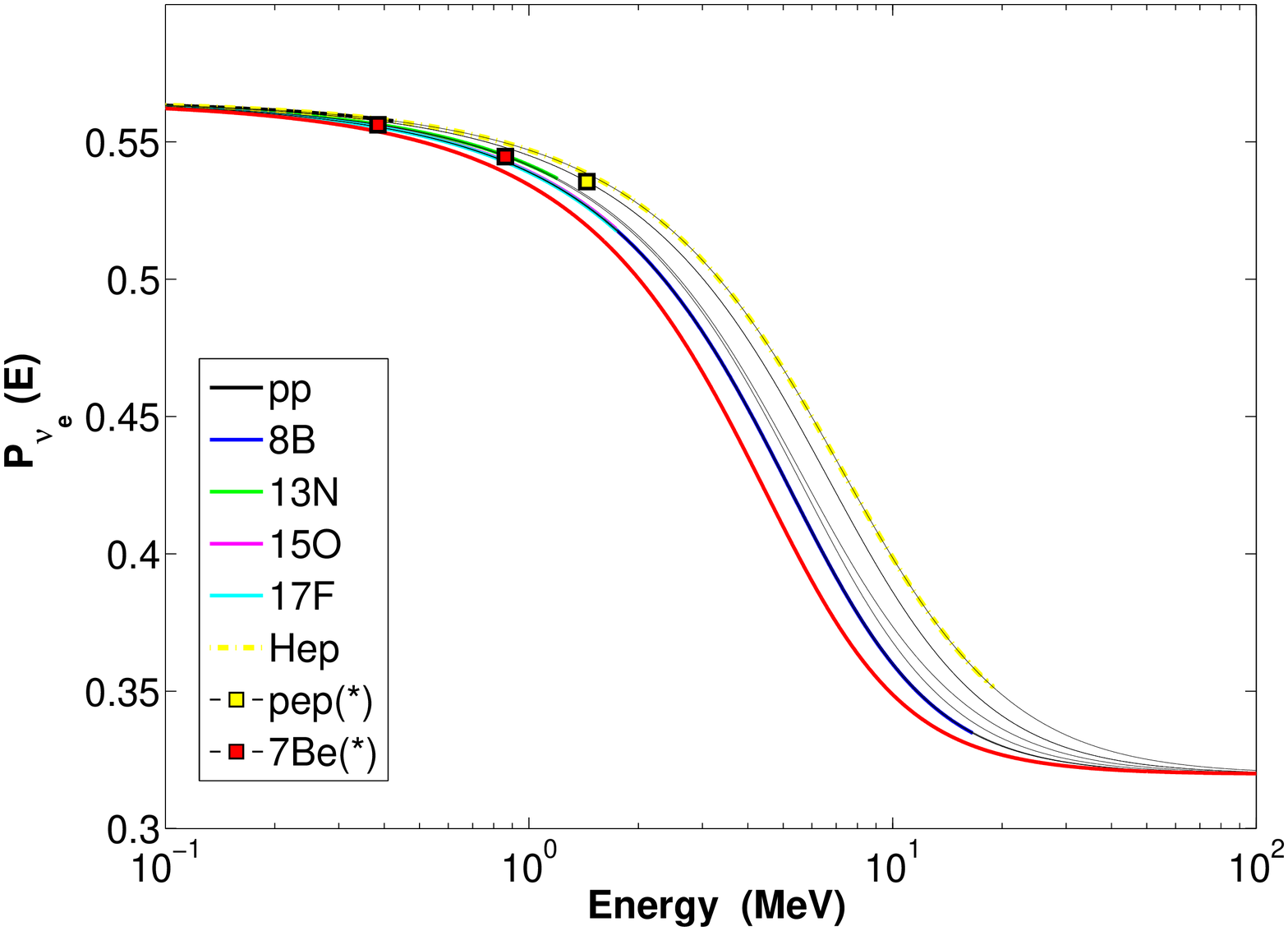}
\includegraphics[scale=0.30]{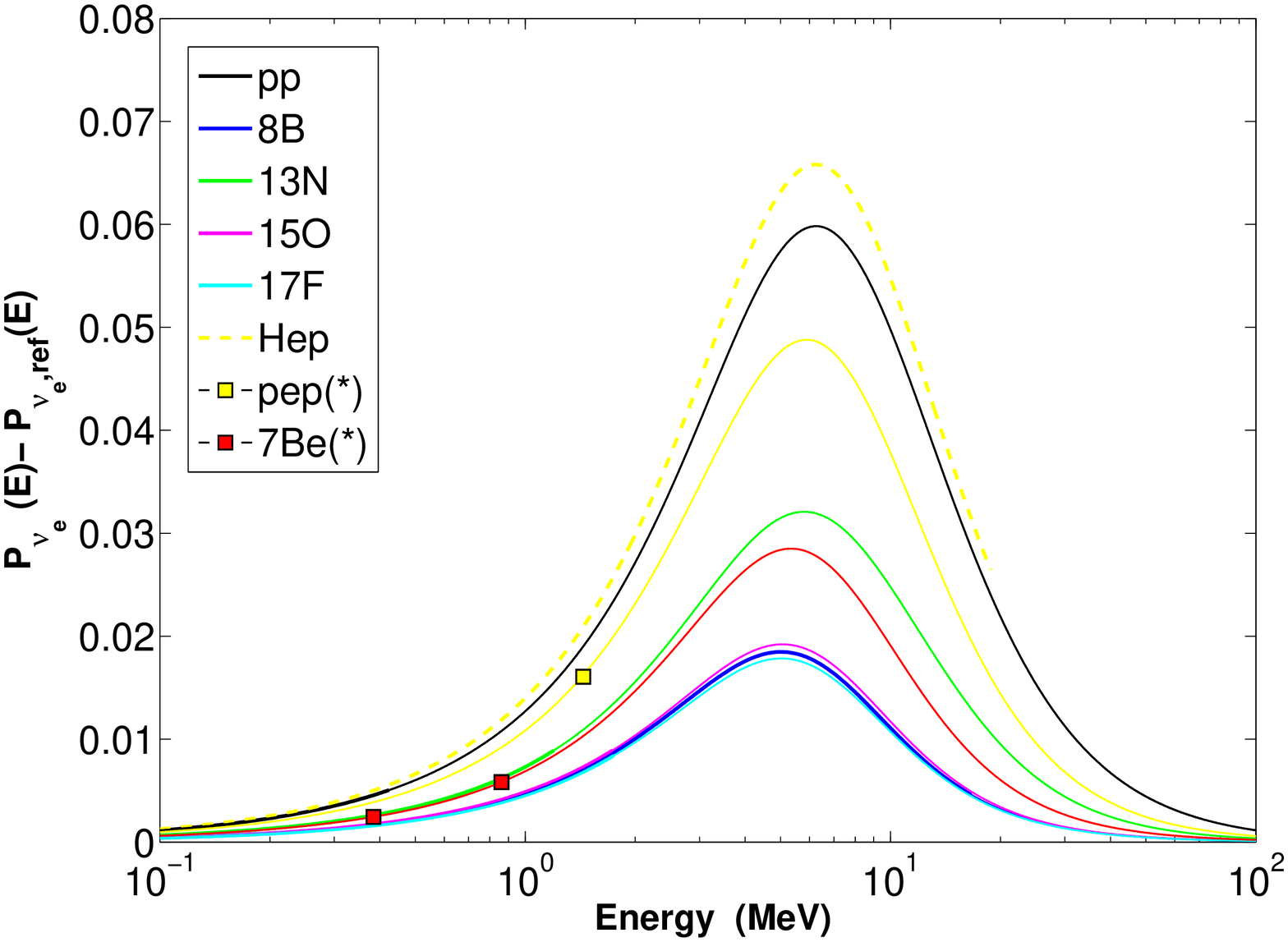}
\caption{The survival probability of electron-neutrinos in function of the neutrino energy 
for the different regions of emission. 
a) survival probability of  electron-neutrinos produced in the different regions 
b) difference between the  survival probability of the different electron-neutrinos and the reference curve corresponding the central emission.
The coloured part of the curves (a), and similarly the bold part of the curves (b), 
indicate the energy range of neutrinos  produced in  the Sun's  core for each nuclear reaction (Cf. Figure \ref{fig:B}).
This reference curve defines the survival probability of electron-neutrinos in the centre of the Sun 
for which the MSW flavour oscillation mechanism is maximum (dashed red curve of figure 3a).}
\label{fig:C}
\end{figure}

Figure~\ref{fig:B} shows the energetic neutrino spectra  \citep{1989neas.book.....B} for an updated version of our standard solar model before and after the flavour conversion.  
Figure~\ref{fig:C}a  shows the survival probability of electron-neutrinos produced in the regions where  occur the different nuclear reactions.
This survival probability of electron-neutrinos  $\langle P_{\nu_e} (E)\rangle_j$
is very similar  for low- and high-energy  neutrinos but presents some differences between.
Unfortunately, as is well known, the emitted neutrino energy spectrum is limited to
a specific energy range for each nuclear reaction. Nevertheless, to highlight the sensitivity of neutrino 
to MSW flavour oscillations, we choose to represent the survival probability of electron-neutrinos in all the available energy range, so that the regions of interest are indicated in colour, or by a single colour square in the case of line fluxes (Cf. Figure 2).

The neutrino change of flavour for low-energy neutrinos is due to vacuum oscillations, 
 and for high-energy neutrinos  it is caused by a cumulative effect of oscillations in vacuum 
 and matter (MSW effect). The experimental neutrino flux measurements for 
 low-energy  neutrinos  have been used to determine the value of $ \theta_{12}$\citep{2010PhRvD..82c3006B}.  
The survival probability  of electron-neutrinos $\langle P_{\nu_e } (E)\rangle_j$  
with intermediate values  of energy,
between 1 MeV and 10 MeV, have a strong dependence on the electron density of the plasma.
Since the production of electron-neutrinos occurs in various nuclear reactions  at 
different layers in the Sun's interior, a sharp differentiation 
is observed between the different $\langle P_{\nu_e} (E)\rangle_j$ curves for the neutrinos with intermediate values of energy.   
 Figure~\ref{fig:C}b shows a well-ordered sequence of curves that corresponds to the difference between the   survival  probability of electron-neutrinos produced by each nuclear reaction $j$
and the survival probability of electron-neutrinos in the  center of the Sun. 
This sequence of curves occurs as a result of the regular decrease of matter density from the center of the Sun.
As a consequence the difference between the survival probability of electron-neutrinos produced by each nuclear reaction $j$ and the survival probability of electron-neutrinos produced in the center of the star increases with the distance of the location of the nuclear reactions to the center (Cf. Figure 3b). For example, in the case of $pp$ electron-neutrinos this difference is of the order of 0.06 which is three times larger than in the case of $^8B$ electron-neutrinos. This is due to the fact that the $^8B$ electron-neutrino source is located near the center of the star (Cf. Figure 1).
Nevertheless, the possibility to observe such effect is somewhat limited, once the neutrino flavour oscillations induced by matter depends on the location of the neutrino source in the Sun's core (cf. Figure 1),  as well as the energy of the emitted electron-neutrinos, i.e., the neutrino emission spectrum (cf. Figure 2). In the case of the $pp$ nuclear reaction, the emitted neutrinos have a maximum energy of 0.42 MeV, consequently  the survival probability difference is smaller than 0.005.  This effect is more expressive in the case of the $^8B$ nuclear reaction for which the emitted neutrinos have a maximum energy of 14.02 MeV. In this case, the effect is maximum for neutrinos with a energy of $4$ MeV  which have a survival probability difference of $0.02$. Identical behaviour will be observed for the CNO nuclear reactions (namely the  nuclear reactions related with the production of $^{13}N$, $^{15}O$ and $^{17}F $ chemical elements) which emitted neutrinos with maximum energies slightly above $1$ MeV, for which the survival probability difference is  of the order of $0.014$. An equally  pronounced difference is observed for the spectral line of $pep$ neutrinos. In the case of the $^7Be$ neutrino the effect is important for the high-energy neutrino line and very small in the case of the small-energy neutrino line. In particular, it should be possible to check experimentally the modification of the  $^8B$ energy neutrino spectrum caused by matter flavour oscillations due to the radial dependence of $^8B$ electron-neutrino source.

\section{The Physics of the Sun's core as probed by neutrinos}

 \begin{table}
 \centerline{\sc Central parameters of standard and modified solar models}
 \centerline{$\qquad$}
 \centerline{
 \begin{tabular}{lccc}
 \hline
 \\
 Solar Models\footnote{\\The  solar standard model (SSM) and the modified models, Sun A and Sun B,   
 were calibrated in order to keep the total luminosity of the present Sun ( Figure~\ref{fig:D} shows the corresponding density profile).}
 &  SSM
  & Sun A & Sun B  \\
  \\
  \hline
 \hline
 {\bf Model Values} &  & &  \\
 density (${\rm g} / {\rm cm}^{3}$)& 151 & 161  & 167 \\
 mean  molecular \\ weight  per electron $\mu_e$ & 1.69 & 1.65 & 1.64 \\
 \\
 \hline
  Neutrino Energy (${\rm MeV}$) 
  \\ Lower Cut-off  ${\rm E_{rc}}$\footnote{
 ${\rm E_{rc}}$  is the minimum  energy that a neutrino must have to be affected by the MSW flavour oscillation mechanism.
 The values between $ [\dots] $ show the percentage difference between the model values 
 and the inverted values. }  
 & 2.0 &1.83 & 1.75  \\ 
 \\
 \hline
 \hline
 {\bf Inverted Values} &  & &  \\
 density (${\rm g} / {\rm cm}^{3}$)\\
 with $\mu_e $ solar model & 151.5 [0.3\%] & 161 [+0.4\%] & 168 [+0.6\%]\\
 density  (${\rm g} / {\rm cm}^{3}$)\\
 with $\mu_e=\mu_{\rm SSM}=1.69$ & 151.5 [0.3\%] & 165 [+2.8\%] & 173 [+3.6\%]\\
 \\
 \hline
 \hline
 \end{tabular}}
 \end{table}
 
The solar model
has been checked  in most of the solar radiative region by
means of the high precision data of SOHO helioseismic instruments \citep{2011RPPh...74h6901T,2012RAA...12h1107T}. 

As shown in equations 1 to 3, the radial profile of the electron density of the solar standard model is a fundamental  ingredient to test the neutrino physics theory. 
 Up to now this quantity has been checked by the detection of the first gravity modes which are really sensitive to the central region of the Sun \citep{2012ApJ...746L..12T,2012RAA...12h1107T}.
In the next few years, with the increase of accuracy  on the measurements 
of several solar neutrino experiments, the situation could reverse:
neutrino fluxes should start to be used to diagnose and to infer the thermodynamic properties of the Sun's core. 

The neutrino flux emissions inside the Sun are sensitive to the local values of  temperature (specifically some of them), 
matter density, chemical composition and  electronic density where the nuclear reactions are taking  place. 

In particular, as the MSW flavour oscillation is strongly dependent on the local electronic density,
it is possible to infer this quantity from  the survival  probability of electron-neutrinos.
Furthermore, since electronic density depends on matter density  and  chemical composition,
it will also be possible to obtain some information about these quantities.

\subsection{The sensitivity of neutrino flavour oscillations to the central density}

Any standard solar model calibrated for the present day total solar luminosity  
has a radial profile of temperature and density in the core that results 
from the balance found between the energy produced by the nuclear reactions 
and the energy transported to the surface. Solar models with slightly different physical assumptions, 
have different  radial profiles of temperature and  matter density among other quantities.
 This is the result of the readjustment of the Sun's internal structure 
caused by the need to obtain the same total luminosity. 

Our  evolution code is an up-to-date version of  the one-dimensional stellar evolution code CESAM
\citep{1997A&AS..124..597M}.  The  code has an up-to-date  and very refined microscopic physics
(updated equation of state, opacities, nuclear reactions rates, 
and an accurate treatment of microscopic diffusion of heavy elements), 
including the solar mixture of \citet{2005ASPC..336...25A,2009ARA&A..47..481A}.
The solar models are calibrated to the present solar radius $R_\odot= 6.9599 \times 10^{10} \;{\rm cm}$, 
luminosity  $L_\odot = 3.846 \times 10^{33} \; {\rm erg\; s}^{-1}$, mass $M_\odot = 1.989 \times 10^{33} \;{\rm g}$, 
and age $t_\odot = 4.54\pm 0.04\; {\rm Gyr}$  \citep[e.g.][]{2011RPPh...74h6901T,2011ApJ...731L..29T}. 
The models are required to have a fixed value of the photospheric ratio $(Z/X)_\odot$, where  
 X and Z are the mass fraction of hydrogen and the mass fraction of elements heavier than helium.
The value of $(Z/X)_\odot$ is determined according to the solar mixture proposed by \citet{2005ASPC..336...25A}.
Our reference model is a solar standard model  that shows 
acoustic seismic diagnostics and solar neutrino fluxes near from other solar standard models 
\citep{1993ApJ...408..347T,turck04,2005ApJ...621L..85B,2010ApJ...713.1108G,2009ApJ...705L.123S,2010ApJ...715.1539T}.

\begin{figure}[!t]
 \centering
 \includegraphics[scale=0.6]{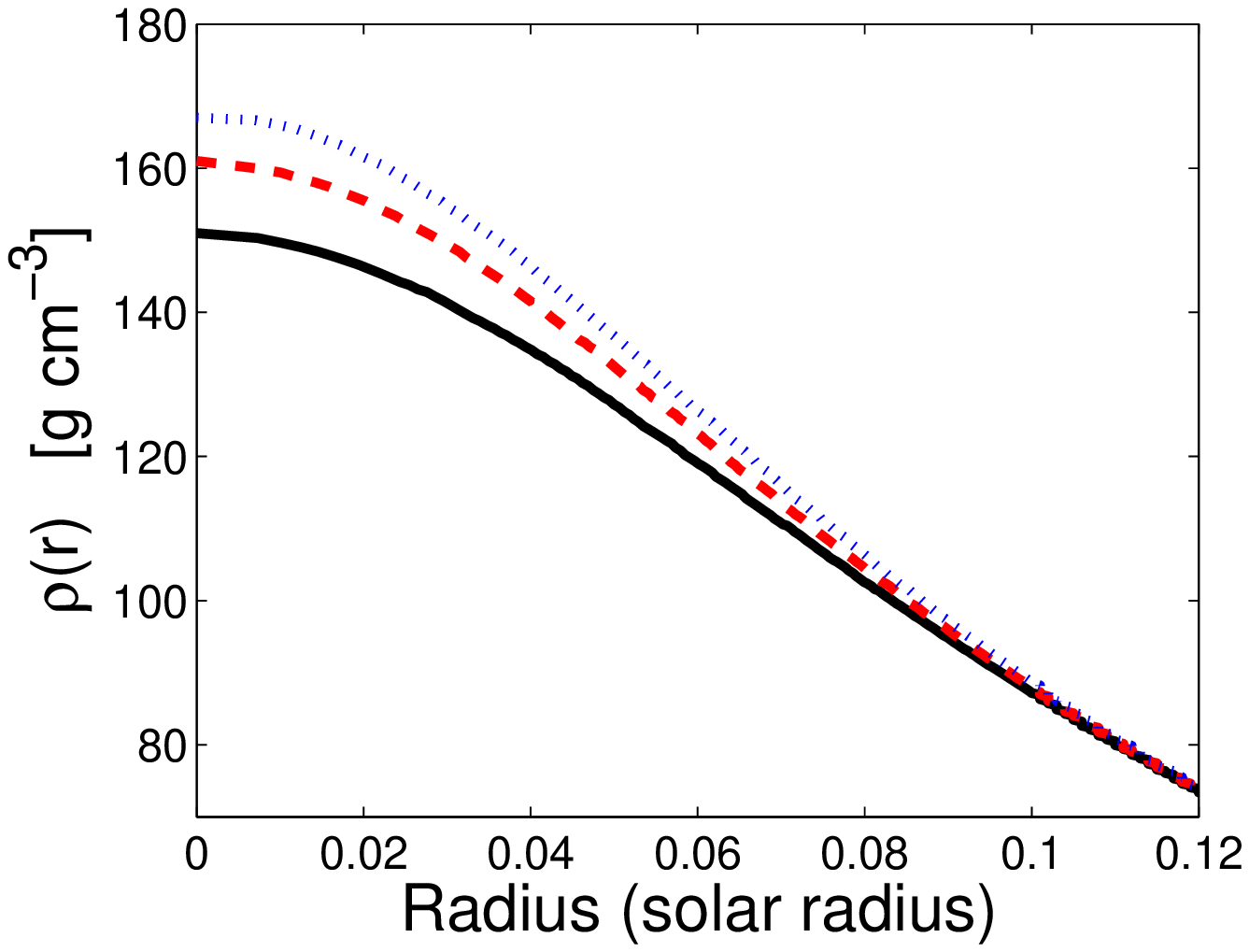}
 \includegraphics[scale=0.6]{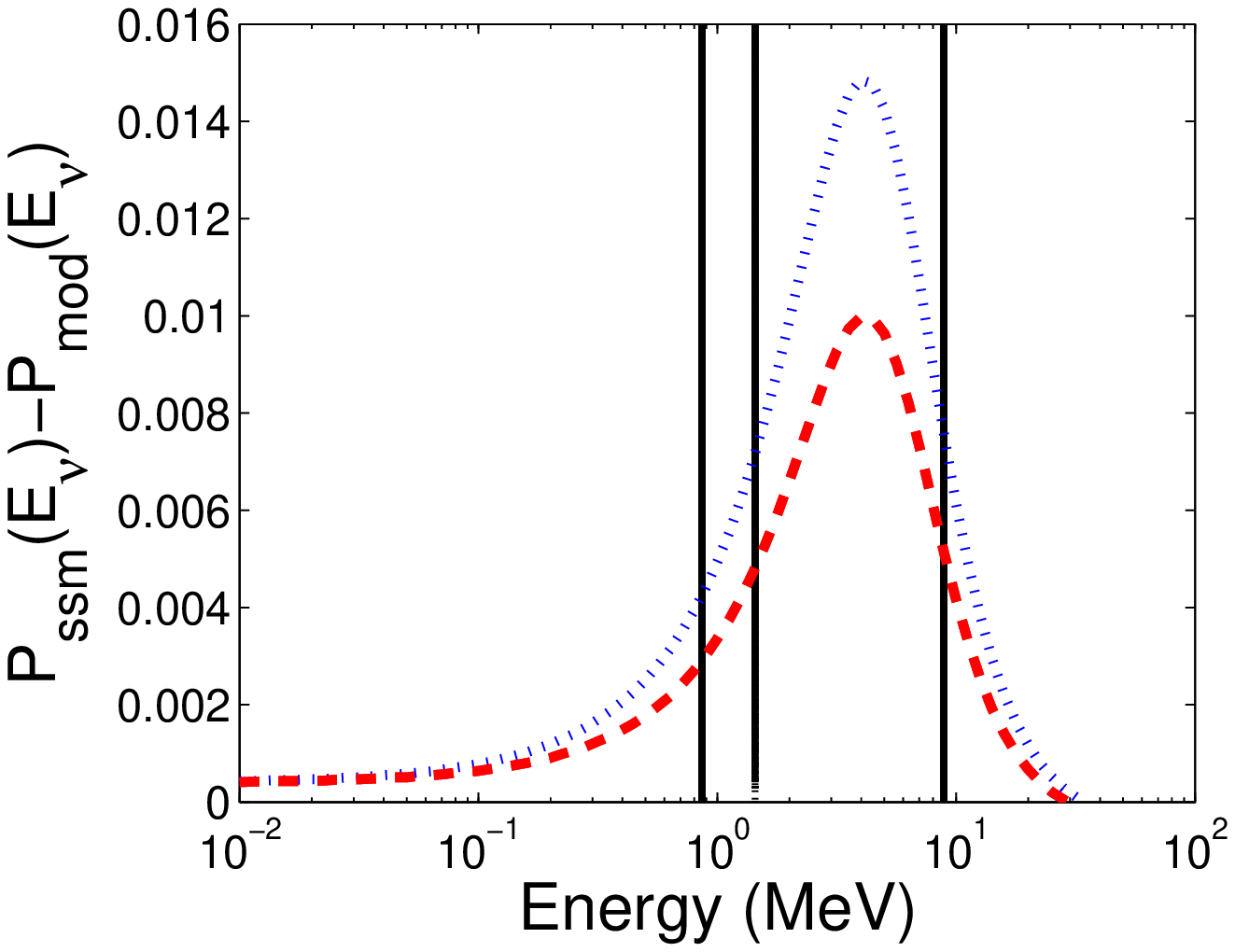}
  \caption{The standard solar model and two modified solar models (see table 1): 
   (a) density as a function of radius in the solar core; 
   (b) difference of the survival probability of electron-neutrinos
  of two modified solar models and the survival probability of electron-neutrinos 
  of the standard solar model. The survival probability of electron-neutrinos
  was computed for neutrinos propagating in the center of the star.
  The continuous-black curve corresponds to the standard solar model, the 
  dashed-red curve  and dotted-blue curve   correspond to the solar models, Sun A and Sun B
  (see table 1).  In the figure, the vertical lines indicate the location of the present solar neutrino measurements: 
   $^7Be$-$\nu$ (0.56 MeV),  $pep$-$\nu$ (1.44 MeV) and $^8B$-$\nu$ (8.9 MeV).}
  \label{fig:D}
\end{figure}

In Table 1 we present the central density and the mean electronic molecular weight of this standard solar model 
and  of two other non-standard solar models.
These last two models have their radiative energy transport
slightly modified in the core in order to obtain 
a Sun's model with a higher central density.  
Figure~\ref{fig:D}a shows the radial density  profile in the Sun's core for the three models.
The amount of neutrinos converted to the non-electron flavour by the MSW flavour oscillation mechanism
is dependent on the central density profile. Figure~\ref{fig:D}b shows the difference 
between the survival probability  of electron-neutrinos for each 
of the two modified solar models relatively to the standard solar model. 
The difference is more significant for the model with the highest central density.
It is clearly illustrated
that an increase of the central density leads to an increase in the neutrinos converted by
the MSW flavour oscillation in all electron-neutrino subspecies. 
   
The survival probability of electron-neutrinos can be used to infer
the matter density in the centre of the Sun, 
once that precise neutrino measurements will become available.
In the following we describe a simple procedure that highlights the
sensitivity of  neutrino flavour oscillations to the 
Sun's core density.
Only neutrinos with energy above a given minimum value
will have their flavour changed by the MSW flavour oscillation mechanism. 
Several resonances occur in  equation (\ref{eq-B}) for neutrinos with energy  $E_r$  
such that    $V(E_r,r) =\Delta m^2 \cos{(2\theta_v)}$.
For the lowest neutrino energy value $E_{\rm r}$ that verifies the previous equation, 
 we choose to call it  $E_{\rm rc}$- the cut-off minimum resonance neutrino energy for a given solar model.
The value of $E_{\rm r}$ is minimum for the maximum 
value of the density of matter (equation \ref{eq-B2}).  
Therefore, each solar model has a unique cut-off minimum 
neutrino energy $E_{\rm rc}$ given by 
\begin{eqnarray} 
 E_{\rm rc}=
 \frac{\Delta m^2 \cos{(2\theta_v)}}{2\sqrt{2} G_f N_o \rho_c /\mu_{ec}}
 \label{eq-D}
\end{eqnarray} 
where $\rho_c$ in the central matter density and
$\mu_{ec}$ is the central value of mean molecular weight per electron.

In Table 1, we show the value of $ E_ {\rm rc} $ for the three solar models computed from
the survival probability electron-neutrino function: if the resonance condition is verified,  
$ \theta_m = \pi/4 $ and the survival probability  of electron-neutrinos
(equation \ref{eq-A}) has the specific value of 0.5. 
The value $E_{rc}$ is determined for each solar model by identifying the neutrino minimum 
energy value  for which  the survival probability of electron-neutrino is equal to 0.5.
The models with the higher central densities have the lowest values of $E_{\rm rc }$
(see Table 1).  The central density  can be calculated from 
equation (\ref{eq-D}) once the value of $ E_ {\rm rc} $ is known for each solar model
and  assuming  that $\mu_{ec}$ (or  the chemical composition) is known.
In Table 1 we show the "inferred values" of  the central density.
The small variation $ \mu_{ec}$ between different models allows us to make a
reasonable estimation of the central density, 
assuming that the variation of resonance energy  $ E_{rc} $ is due 
solely to variations in the central density. 
The central density is inverted at most
with a difference of a few percent higher than the value of the model, 
even in the case that $ \mu_{ec}$ is poorly known. 

In fact, the central value of the density in the Sun might be more difficult
to obtain  since the electron-neutrinos are not produced just in the center,
and the different ranges of  energy of neutrinos are limited 
(Cf. Figure~\ref{fig:B}). An interesting possibility  is to
probe the solar core by using the neutrinos produced 
near the centre of the star, such as $^8B$-$\nu$,  $^{15}O$-$\nu$ and $^{17}F $-$\nu$ neutrino fluxes.
The simultaneous measurement of  two different neutrino fluxes 
produced in the same layer of the Sun's interior, such as
$^8B$-$\nu$ and  $^{15}O$-$\nu$, 
could be used to estimate simultaneously the matter density 
and the average molecular weight per electron 
in the core of the Sun.

\subsection{The sensitivity of neutrino flavour oscillations to the radial density profile}

All the solar models validated by  acoustic and gravity mode detections, 
predict that the density of matter decreases by 90\% 
from its central value to the value at 30\% R$\odot$.
The neutrino fluxes produced by the different nuclear reactions are sensitive 
to the local values of the electronic density. 
The different neutrino fluxes $\Phi_j(r)$ can be interpreted as a local average of $\rm n_j (r)$\footnote{$\rm  n_ {j} (r)$ 
is the estimated mean value of the electronic density for the layer at the radius $\rm r_j$,
where the neutrino source $\Phi_j(r)$ is maximum  (Cf. figure~\ref{fig:A}).}.

 In principle taking into account the high sensitivity of solar neutrino fluxes to the temperature changes in the solar core, we should expect important variations in the location of the neutrino source in the Sun's core. However, the effect is actually quite small. The reason is related with the calibration procedure that solar models are subjected to, which require that the Sun in the present age must have the observed solar luminosity. Consequently, the production of neutrinos inside the Sun, which are strongly dependent on the luminosity of the star, namely, the neutrinos produced in the proton-proton chain nuclear reactions, such as $^8B$ and $^7Be$,  occur sensibly in the same location,  i.e., at the same distance from the center of the star.  As an example, solar models which experiment a decrease of 10\% in the central density  (and a 3\% reduction in the central temperature) have a variation of the maximum of any of the neutrino sources (both proton-proton chain nuclear reactions and CNO nuclear reactions) displaced by an amount smaller than 1\% of the solar radius, although the neutrino fluxes change significantly due to their high sensitivity to the temperature.  Actually, the proton-proton neutrino source has the same location, as this nuclear reaction, more than any other depends directly from the total luminosity of the star. For the purpose of this article the neutrino sources of different solar models are considered to be located within the same fraction of the solar radius. Therefore,  if the fundamental parameters of neutrino oscillations are known, the  probability of survival of electron neutrinos can be used to measure  the radial electronic density profile in the solar core, i.e., the matter density and molecular weight per  electron in several layers of the Sun. A simple  procedure to obtain the electronic density 
is now proposed.  We consider first an electronic-neutrino survival probability $\bar{p}_j$  determined,  
at a specific neutrino energy $E_j$, 
by some neutrino
experiment. One assumes that $\bar{p}_j$ is determined fully independently from
the solar standard model by a method identical to the one described in~\citet{2001PhLB..521..287B}.
Preferentially,  $E_j$  
is the value chosen within the energy interval for which the MSW neutrino flavour 
oscillations are significant (cf. Figure~\ref{fig:C}).  
For each pair  $(E_j, \bar{p}_j)$, the value of the electronic density for a certain layer 
can be computed from equations~(\ref{eq-A}-\ref{eq-B2}).
We estimate $ \bar{n}_ {j} \equiv \left\langle n_e (r)\right\rangle_j$ corresponding to electronic density
of several layers in the Sun's core, as
$ \bar{n}_ {j}= 1/(2 \sqrt {2} G_f) \times \alpha_j  \left\langle V\right\rangle_j / E_ {j} $,  
where  $\left\langle V\right\rangle_j $ is computed from $p_j$ using the previous equations,
and $\alpha_j $   is a weight-correction parameter unique  for each $j$-nuclear reaction.  
$\alpha_j $ is estimated for the solar standard model and it takes a value of the order of a unit,
increasing slightly when the neutrino source $\Phi_j(r)$  moves away from the centre of the star.
Similarly, by using the experimental value $\bar{p}_j$, which is different from 
the theoretical prediction  $p_j$, we estimate the value $\bar{n}_j$ 
as a correction to the theoretical value  $n_j$, from the equation 
$ \Delta n_{j}/ n_{j} =  \beta_{j}\; \Delta p_{j}/ p_{j}$, 
where  $\Delta n_{j}= \bar{n}_{j}-n_{j} $, $\Delta p_{j}=\bar{p}_{j}-p_{j}$ and
$\beta_{j}$ is a coefficient computed from the solar standard model.
The previous expression is  obtained from a perturbation analysis 
of equations~(\ref{eq-A}-\ref{eq-B2}).
In this analysis  $\bar{n}_j$  is the inverted value deduced from the experimental data.

\begin{figure}[!t]
\centering
\includegraphics[scale=0.40]{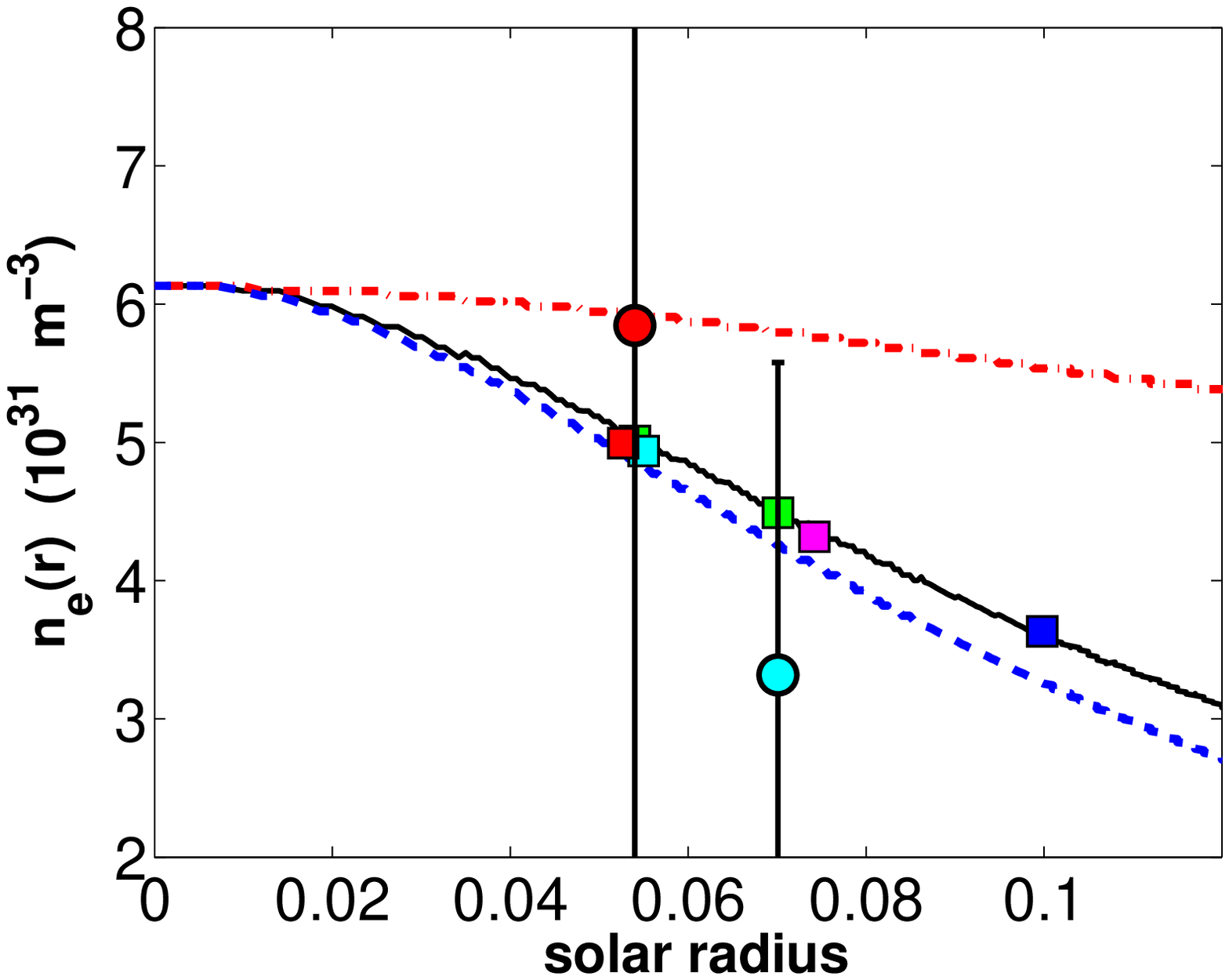} 
\includegraphics[scale=0.40]{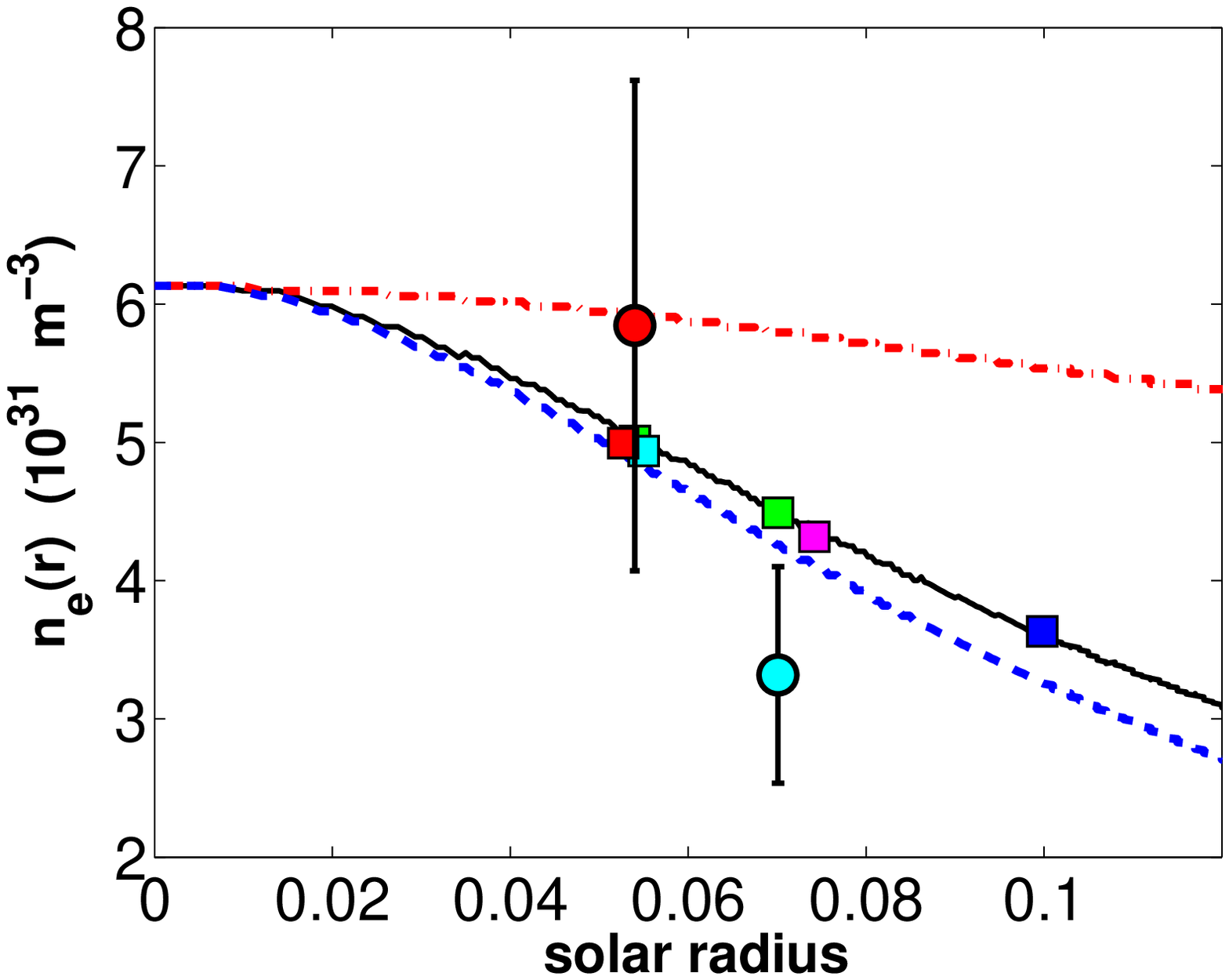} 
\caption{
Radial profile of the electronic density $\rm n_e (r) $ (black continuous curve) in the core of the Sun of the  solar standard model,
on the same plot appear normalized values of
the radial profile of the density of matter, central $\rho$ = 167 ${\rm g\; cm^{-3}}$ (blue dashed curve) 
and  the radial profile of molecular weight per electron $\mu_e(r)$,  central value of 1.69 (red dotted curve). 
The points shown in the figure correspond to the values  of density  {\it inverted} from
the  survival probability of electron-neutrinos, at the following energies: 
0.56 MeV ( $^7Be$),    1.44 MeV ( $ pep$),    8.9   MeV ($^8B$),
1.0  MeV  ($^{13}N$), 1.0   MeV ( $^{15}O$)    1.0  MeV ($^{17}F$).
The location of the source are the following ones:
0.070 $R_\odot$ ($^7Be$, square-green),   0.099  $R_\odot$ ($ PeP$, square-blue), 
0.054  $R_\odot$ ($^8B$,  square-green),0.074  $R_\odot$  ($^{13}N$, square-magenta),  
0.055 $R_\odot$ ( $^{15}O$, square-cyan) , 0.053  $R_\odot$ ($^{17}F$, square-red).
The inverted electronic density computed from measured $^8B$ and $^7Be$  
neutrino fluxes: (a) the values 
computed from the survival neutrino probability obtained from the present results
of SNO ($^8B$) and Borexino ($^7Be$),
(b) idem considering that the experimental error 
in the determination of the survival probability of electron neutrinos is 4 \%.
The red and in cyan circles correspond to the $^8B$ and $^7Be$  neutrino measurements
respectively.
}
\label{fig:E}
\end{figure}

Figure~\ref{fig:E} shows the inverted electronic density values obtained from the 
survival probability of electron-neutrinos as described previously.  There is a good 
agreement between  the values of the electronic  density obtained by inversion 
and  the electronic density of the standard solar model.  
On Figure~\ref{fig:E}a we show the electronic density values that  one can deduce from the survival probabilities obtained from the neutrino detector measurements. 
More specifically,  we present the electronic density inverted 
from the  $^7Be$ neutrino fluxes  using the measurements of~\citet{2010PhRvD..82c3006B} and  the   $^8B$ neutrino fluxes  using the measurements of~\citet{2007PhRvC..75d5502A}.

The Borexino experiment  measures $^7Be$ solar neutrino rates with an 
accuracy better than $5\%$. This corresponds to a $^7Be$  neutrino flux  
$\Phi(^7Be)= 4.87^{+0.24}_{-0.24}  \times 10^9\; {\rm cm^{-2} s^{-1}}$,
under the assumption of the MSW-LMA scenario of solar neutrino oscillations
\citep{2011PhRvL.107n1302B,2008PhRvL.101i1302A}. The estimated survival probability, assuming a high metallicity solar standard model, 
was initially  estimated to be
$P_{\nu_e} (^7Be)=0.56\pm 0.1$ at the energy $0.862 \; {\rm MeV}$ 
\citep{2008PhRvL.101i1302A}. Latter the Borexino collaboration\citep{2011PhRvL.107n1302B} 
updated their estimation to $P_{\nu_e} (^7Be)=0.52^{+0.07}_{-0.06}$.
Similarly, as proposed by~\citet{2001PhLB..521..287B}, we compute the survival probability 
for $^8B$ as $P_{\nu_e}(^8B)=\Phi_{\nu_e}(^8B)/\Phi_{t}(^8B)$,
where $\Phi_{t}(^8B)$ is the total neutrino flux integrating all neutrino flavours measured by SNO  \citep{2010PhRvC..81e5504A}:
$\Phi_{t}(^8B)=5.046^{+0.226}_{-0.275}  \times  10^6\; {\rm cm^{-2} s^{-1}}$
and $\Phi_{\nu_e}(^8B)$  is the electron-neutrino flux measured by the same detector \citep{2007PhRvC..75d5502A}: 
$\Phi_{\nu_e}(^8B)=1.76^{+0.14}_{-0.14}  \times  10^6\; {\rm cm^{-2} s^{-1}}$. It follows that $P_{\nu_e}(^8B)=0.35\pm 0.047$ for SNO.
It is clear that presently the error bars are still too large to really estimate the small gap with the standard model values. But one may hope a progress on the future as we have already for $^8B$ neutrinos several improved detections: the neutrino flux measured by the Kamiokande-III experiment \citep{2011PhRvD..83e2010A}, 
$\Phi_{\nu_e}(^8B)=2.32^{+0.09}_{-0.09}  \times  10^6\; {\rm cm^{-2} s^{-1}}$,
the neutrino flux measured by the Borexino experiment
is $\Phi_{\nu_e}(^8B)=2.4^{+0.5}_{-0.5} \times  10^6\; {\rm cm^{-2} s^{-1}}$ \citep{2010PhRvD..82c3006B}.
Moreover, we already measure the energy dependence of these fluxes down to 3 MeV.
So, to illustrate the potential of this diagnostic and to indicate some research perspectives,  we present in  Figure~\ref{fig:E}b the electronic density values deduced assuming an error bar on the survival probability of the order of 4\%. It follows that the error bar on the electronic density that one can deduce for $^7Be$ and $^8B$ fluxes is largely reduced. This is due to the high sensitivity of neutrino flavour oscillations to the electronic density. 
This first study shows the potentiality  of such type of neutrino diagnostic.
Another possibility is to fix values of the electron-neutrino survival probability curve 
for  neutrinos of low and high energy, as such neutrinos have respectively 
pure vacuum oscillations and vacuum plus MSW oscillations. 
In this case neutrinos with energy in the interval $0.5-10$ MeV, can probe
the electronic density  of different solar layers. 
 
\section{Discussion and Conclusion} 

The capability to make such a study is real because  new neutrino detection experiments are arriving and others are planned for the near future. The diversity of neutrino flux measurements expected at different energy levels will allow us to make reliable and detailed neutrino inversions.

In this first study we have used only two mean survival probabilities, without taking into account the energy dependence of the boron flux and the detailed radial distribution. But the Borexino detector also measures the  $pep$ neutrino flux, 
$\Phi(pep)= 1.6^{+0.3}_{+0.3} \times 10^8 {\rm cm^{-2} s^{-1}}$ \citep{2012PhRvL.108e1302B}.
This preliminary result is still insufficiently accurate, but it is not in contradiction with the fact that the Sun is hotter than what is suggested by the present standard solar model. 
The $pep$  neutrino flux is strongly dependent on the luminosity of the star. 
Therefore, it is an indirect measurement of the total energy (luminosity)
produced in the nuclear region. So it represents  a powerful probe to the physics of the nuclear region of the Sun, in parallel to the information introduced in the seismic model \citep{2012RAA...12h1107T}.

Today, there is some uncertainty about the chemical composition in the Sun's core.
In parallel, all the neutrino measurements agree with the predicted values of  the solar seismic  
model but not so well with the standard model \citep{2011RPPh...74h6901T}, showing 
some capability of understanding the difference with the standard model if one extracts the electronic density profile.
The results obtained so far suggest that
in the near future we will be able to probe the physics of the Sun's core using neutrinos.

In this work we have proposed a strategy that allows to use the solar neutrino flux measurements to invert the 
electron density of the solar plasma at specific locations of the Sun's core.
Furthermore, in the near future, it will also be possible to put important constraints on the matter density
and on the molecular weight per electron of the solar plasma. 
The inversion is made based upon the assumption that the future Earth neutrino experiments 
 (e.g., reactor experiments, superbeams, beta beams and neutrino factories) 
will allow the precise determination of basic neutrino oscillation parameters.
  
Today,  the neutrino fluxes of  $^7Be$-$\nu$,  $^8B$ -$\nu$ and $pp$-$\nu$ 
are already well measured and obtained separately. The  $pep$-$\nu$ neutrino
flux is also measured although with much less precision. It will be
interesting to verify if such results hold with the increase of accuracy in the observations within the
next few years. The present electronic-neutrino survival probability curve
is fixed at low neutrino energy by $pp$-$\nu$, sensitive to pure vacuum oscillations,
and by  $^8B$-$\nu$, sensitive to vacuum plus MSW oscillations. Therefore it will
be relevant to verify if future measurements of $^7Be$-$\nu$ and 
$pep$-$\nu$ can give us some information about the electronic density 
in two distinct layers of the Sun, namely  $0.1\; R_\odot$ ($pep$-$\nu$) and $0.07\; R_\odot$ ($^7Be$-$\nu$).
There is a real possibility that Borexino or SNO + experiments 
will be able to measure accurately these different neutrino fluxes, 
as well as the neutrino fluxes of some of the CNO cycle neutrino emission
reactions. 

The LENA (Low Energy Neutrino Astrophysics) solar neutrino detector 
is expected to start to operate within the next few years.  
This detector will be able to perform very accurate  measurements of the different 
sources of the solar neutrino spectrum, 
allowing not  only very precise measurements of the solar core plasma, 
but also to identify  possible seasonal neutrino variations. 

In conclusion, we have shown that in the near future one may hope to use neutrino spectroscopic measurements to infer the electronic density of the plasma in the core of the Sun.  This will be an important and totally independent test  of the nuclear region of the Sun properties in parallel with the helioseismic inversions of matter density and sound speed. The proposed method, that could be improved, allows to infer the properties of the plasma in the very central region of the Sun, which until now have only been explored by some gravity mode detections with GOLF/SoHO. We have shown in table 1 that the inversion of different density profile can be done with a reasonable quality if the neutrino fluxes are obtained with a good accuracy. Of course this method supposes that the neutrino oscillations description is independent of solar model predictions. If that becomes the case, one will be able to extract from the different solar neutrino detections, the electronic density and matter density profiles with some hints on the CNO composition in the core of the Sun.

\begin{acknowledgments}
This work was supported by grants from "Funda\c c\~ao para a Ci\^encia e Tecnologia"  and "Funda\c c\~ao Calouste Gulbenkian".  We thank the anonymous referee for the helpful comments that much improved
the clarity of the paper. 
\end{acknowledgments}


 
 

\end{document}